\documentclass[%
 reprint,
 amsmath,amssymb,
 aps,
]{revtex4-2}

\usepackage{graphicx}
\usepackage{dcolumn}
\usepackage{bm}
\usepackage{stmaryrd}

\makeatletter
\DeclareRobustCommand{\cev}[1]{%
  {\mathpalette\do@cev{#1}}%
}
\newcommand{\do@cev}[2]{%
  \vbox{\offinterlineskip
    \sbox\z@{$\m@th#1 x$}%
    \ialign{##\cr
      \hidewidth\reflectbox{$\m@th#1\vec{}\mkern4mu$}\hidewidth\cr
      \noalign{\kern-\ht\z@}
      $\m@th#1#2$\cr
    }%
  }%
}
\makeatother

\begin{document}


\title{Quantum information in Riemannian spaces}

\author{Pablo G. C\'amara}
 \email{pcamara@upenn.edu}
\affiliation{%
 University of Pennsylvania, 3700 Hamilton Walk, Philadelphia, PA 19104, USA
}%




\date{\today}

\begin{abstract}
We present a diffeomorphism-invariant formulation of differential entropy for Riemannian spaces, providing a fine-grained, coordinate-independent notion of quantum information for continuous variables in physical space. To this end, we consider the generalization of the Wigner quasiprobability density function to arbitrary Riemannian manifolds and analytically continue Shannon's differential entropy to account for contributions from intermediate virtual quantum states. We illustrate the framework by computing the quantum phase-space entropy of harmonic oscillator energy eigenstates in both Minkowski and anti-de Sitter geometries. Furthermore, we derive a generalized entropic uncertainty relation, extending the Bialynicki-Birula and Mycielski inequality to curved backgrounds. By bridging concepts from information theory, differential geometry, and quantum physics, our work provides a systematic approach to studying continuous-variable quantum information in curved spaces. 
\end{abstract}

\maketitle


\section{Introduction}

Over the past century, the concept of information has emerged as a central idea across nearly every corner of science, deeply influencing disciplines such as physics, biology, computer science, and electrical engineering. Information can be understood as constraints within the space of possible outcomes of a measurement, known as \emph{sample space}, and it serves as the foundation for computation. When the elements of the sample space are countable, this notion can be precisely formalized through Shannon's entropy \cite{shannon1948mathematical} (or its counterparts, Gibbs and von Neumann entropies in statistical and quantum mechanics, respectively), which quantifies the amount of uncertainty associated with the system prior to the measurement. However, there are many instances in nature in which the sample space is given by a continuum of possible outcomes endowed with a metric structure. Physical particles, for example, can move within continuous spatial dimensions and, if unconfined by a potential, can do so with continuous momentum. Relatedly, living organisms have evolved to store information in the possible 3-dimensional conformations that a protein can adopt through allosteric changes, the 2-dimensional electric potentials across neuronal membranes, or the concentrations of distinct species of molecules in cellular compartments \cite{phillips2012physical}. In computer science, the differentiable parameters of a data-generating model span a statistical manifold with a metric given by Fisher information \cite{amari2016information}. In each of these situations, information is described by a probability density function over a Riemannian sample space. 

Shannon's entropy can be extended to uncountably infinite sample spaces by replacing the sum over discrete states with an integral over the sample space, resulting in the concept of differential entropy \cite{shannon1948mathematical}. However, this extension introduces certain complications. For the total probability of an event to remain invariant under coordinate transformations of the sample space, the probability density function must transform inversely to the Jacobian of the transformation. As a result, the differential entropy is not invariant under coordinate transformations of the sample space. In practice, this complication is addressed by considering the relative entropy, or Kullback-Leibler divergence, between probability density functions, which remains invariant under coordinate transformations \cite{pawitan2001all}. Nonetheless, this raises a fundamental problem from a physical standpoint, as it implies the absence of a microscopic notion of information that is independent of the observer's choice of coordinates. If information has an underlying objective reality, there must be a mathematical formulation of entropy for Riemannian sample spaces that is invariant under diffeomorphisms and consistent with the principles of quantum physics and special relativity.

In this work, we initiate the development of such an invariant formulation. Our main result is the quantum generalization of the classical diffeomorphism-invariant differential entropy for Riemannian spaces introduced in \cite{pennec2006intrinsic}, along with the incorporation of momentum degrees of freedom. The resulting entropy extends the quantum phase-space entropy proposed by Laguna and Sagar \cite{laguna2010shannon, salazar2023phase} and by Cerf and collaborators \cite{hertz2017entropy, van2021quantum, cerf2024complex} to arbitrary Riemannian geometries.

The study of relativistic aspects of quantum information is a highly active field of research, as reviewed e.g. in \cite{peres2004quantum, mann2012relativistic}. Much of the work in this area focuses on understanding the relationship between quantum measurements performed by observers in different reference frames. Here, we consider the simplest case of a complete system - without specifying a measuring apparatus or environment - where all microscopic degrees of freedom are accessible. While this is an idealized scenario, a detailed understanding of it is essential for addressing the relativistic quantum measurement problem in practical settings like quantum communication channels. For instance, Peres \emph{et al.} \cite{peres2002quantum} showed that the spin entropy of a free spin-$\frac12$ particle is not a relativistic scalar and lacks an invariant meaning independent of the observer. This non-invariance arises from entanglement between spin and momentum degrees of freedom, and it was argued that the entropy of the complete system, where momentum degrees of freedom are not integrated out, should exhibit invariance \cite{peres2002quantum}. 

In developing a diffeomorphism-invariant formulation of entropy that incorporates position and momentum degrees of freedom, we are led to extend the Wigner function to curved geometries. While multiple studies have generalized the Wigner function to highly symmetric spaces, such as Lie groups and spaces of constant curvature (e.g. \cite{agarwal1981relation, varilly1989moyal, figueroa1990moyal, brif1998general, brif1999phase, alonso2002wigner, alonso2003wigner, mukunda2004wigner, mukunda2005wigner, klimov2008generalized}), these formulations rely on the specific symmetries of those spaces and are not invariant under general canonical transformations. More recently, Gneiting \emph{et al.} \cite{gneiting2013quantum} introduced an expression for the Wigner function that is valid in arbitrary curved geometries and invariant under canonical transformations. We became aware of this work during the revision of our manuscript. Our expression for the Wigner function is mathematically equivalent to that of Gneiting \emph{et al.}, although our derivation is based on the relationship among the Riemannian, Lebesgue, and symplectic measures. 

In this paper, we focus on stationary states and diffeomorphism invariance, while leaving relativistic effects and Lorentzian invariance to future work. The structure of the paper is as follows. In Section \ref{sec1}, we revisit the mathematical formalism of \cite{pennec2006intrinsic} from a physical perspective. Section \ref{sec2} extends this framework to account for momentum degrees of freedom, introducing the concept of phase-space entropy, which corresponds to Gibbs entropy at the classical level. In Section \ref{sec3}, we consider deformation quantization and generalize the Wigner quasiprobability density functions to arbitrary Riemannian spaces. In Section \ref{sec4}, we generalize Shannon's entropy formula to quasiprobability distributions. Equipped with these tools, we conclude Section \ref{sec4} with a precise invariant formulation of phase-space entropy for quantum states in arbitrary Riemannian spaces. In Section \ref{sec5}, we illustrate the computation of quantum phase-space entropy using the non-relativistic harmonic oscillator in both Minkowski and anti-de Sitter spacetimes. Section \ref{secbounds} then explores several aspects of quantum entropic information bounds. Specifically, we generalize the Bialynicki-Birula and Mycielski (BBM) inequality \cite{bialynicki1975uncertainty} to arbitrary Riemannian spaces and discuss violations of the more stringent inequality proposed by \cite{hertz2017entropy, van2021quantum} due to the effects of spatial curvature. Finally, we conclude in Section \ref{discus} with a discussion of the broader implications of our results.

\subsection{Notation}

Throughout the paper, we use the following notation. Probability density functions with respect to the Riemannian measure are denoted by $\hat{\rho}$. Similarly, probability density functions with respect to the Lebesgue or symplectic measures are denoted by $\rho$. Quantum operators are denoted by bolded upper case letters (e.g., $\mathbf{X}$, $\mathbf{P}$, $\ldots$). The exponential chart at a point $x$ in a $D$-dimensional Riemannian space $X$ is denoted by $(U_x,\,\varphi_x)$ and is given by $(X-C_x,\,\log_x)$, where $C_x\subset X$ is the cut locus of $x$, and $\log_x(p)={x^\mu(p)}$ is the diffeomorphism from $U_x$ onto the open subset $\mathcal{D}_x\subseteq T_xX\simeq \mathbb{R}^D$ given by the inverse of the exponential map at $x$. The representation of a function, tensor, or operator $f$ in the exponential chart at $x\in X$ is shorthanded as $f_x(\vec{x})\equiv f(\exp_x(\vec{x}))$. Since the cut locus has zero measure, integrals over a connected and geodesically complete Riemannian space $X$ can be equivalently expressed as integrals over $\mathcal{D}_x$ for any chosen exponential chart $(U_x,\,\varphi_x)$ \cite{chavel1995riemannian}. We assume the metric tensor in any exponential chart has intrinsic dimension of $\textrm{(length)}^2$.

\section{Differential entropy in measurable metric spaces}
\label{sec1}

We consider an observer, Alice, who lives in a connected and geodesically complete $D$-dimensional Riemannian space $X$ and performs measurements of a distant object also located in $X$. Without loss of generality, we assume she does this by measuring the force exerted by the object on a test particle in her laboratory through some interaction. This process requires Alice to track changes in the position of the test particle over time. At a given time $t_0$, such an arrangement can be described in terms of a measurable metric space $\mathcal{X} = (X, d_X, \mu)$, where $d_X$ denotes the geodesic distance in $X$ and $\mu$ is the probability measure associated with the positional observations of the test particle by Alice.  

In this scenario, two different measures are defined in $X$. The first is the probability $\mu(A)$ that quantifies the likelihood (in a frequentist sense) of the test particle being in a specific patch $A\subseteq X$ at $t_0$. The second is the Riemannian measure $\nu_X$ induced by the volume form $dX$ on the Borel sets of $X$,
\begin{equation*}
\nu(A) = \int_{A\subseteq X}dX = \frac{1}{l_{P}^D}\int_{\log_x(A)}\sqrt{\det(g_x(\vec{x}))}\, d^D x
\end{equation*}
where the latter integral is calculated over an exponential chart $(U_x, \varphi_x)$, and $g_x(\vec{x})$ is the (positive-definite) metric tensor of $X$ expressed in the local coordinates. For convenience, we have chosen $dX$ to be dimensionless by using the fundamental length scale of the underlying microscopic theory, which here we assume to be the Planck length $l_P$. However, the resulting phase-space entropy will be independent of the particular normalization choice.

Since $\mu$ is absolutely continuous with respect to $\nu$, namely $\nu(A)=0\implies \mu(A)=0$, by the Radon-Nikodym theorem \cite{billingsley2017probability} there is a probability density function $\hat{\rho}$ with support on $X$ associated to $\mu$
\begin{equation}
\mu(A)=\int_{A\subseteq X} \hat{\rho}\, dX = \frac{1}{l_P^D}\int_{\log_x(A)} \hat{\rho}_x(\vec{x})\sqrt{\det(g_x(\vec{x}))}\, d^D x \label{mua}
\end{equation}

With these definitions, $\hat{\rho}_x(\vec{x})$ is dimensionless and invariant under reparametrizations of the local coordinates, $\hat{\rho}_y(\vec{y})=\hat{\rho}_x(\vec{x}(\vec{y}))$. This contrasts with the usual transformation properties of probability density functions, which are typically defined with respect to the Lebesgue measure \cite{billingsley2017probability, pawitan2001all}. Thus, it is customary for Alice to define the following probability density function in her laboratory with respect to the local Lebesgue measure
\begin{equation}
\rho_{x}(\vec{x})\equiv \frac{1}{l_P^D}\,\hat{\rho}_x(\vec{x})\sqrt{\det(g_x(\vec{x}))} \label{phat}
\end{equation}
where $\{x^\mu\}$ denotes the coordinates in Alice's chart, $(U_x,\varphi_x)$, and the probability density function is normalized in $\mathcal{D}_x$. In particular, $\rho_{x}(\vec{x})$ has dimensions of $\textrm{(length)}^{-D}$ and transforms in the standard way under orientation-preserving reparametrizations of the local coordinates, $\{x^\mu\}\to\{y^\mu\}$,
\begin{equation}
\rho_{y}(\vec{y})=\rho_{x}(\vec{x}(\vec{y}))\, J_{\vec{x}\to\vec{y}}(\vec{y})\label{tphat}
\end{equation}
where $J_{\vec{x}\to\vec{y}}(\vec{y})\equiv\det\left(\frac{\partial x^\mu(\vec{y})}{\partial y^\nu}\right)$ is the Jacobian determinant of the coordinate transformation.

Let us now consider the information gain associated with Alice's observations, given by the differential entropy of $\hat{\rho}$ \cite{pennec2006intrinsic},
\begin{multline}
H_{\mathcal{X}}\equiv -\int_X \hat{\rho}\, \textrm{log}(\hat{\rho})\, dX \\
= -\frac{1}{l_P^D}\int_{\mathcal{D}_x} \hat{\rho}_x(\vec x)\, \textrm{log}(\hat{\rho}_x(\vec x)) \sqrt{\det(g_x(\vec{x}))}\, d^Dx \label{hx}
\end{multline}
This expression is invariant under reparametrizations of the local coordinates and, therefore, independent of the observer. Since $H_{\mathcal{X}}$ depends on both measures, $\mu$ and $\nu$, changes in the geometry of $X$ mediated by gravitational interactions may affect the amount of information associated with Alice's observations. For example, when $X$ is compact and $\hat{\rho}$ is the uniform distribution, the differential entropy $H_{\mathcal{X}}$ is fully determined by the volume of $X$, with $H_{\mathcal{X}}=\textrm{log}(\nu(X))$, and the problem of entropy estimation is equivalent to that of volume estimation. 

In contrast, the differential entropy that Alice would conventionally define in her laboratory using the Lebesgue measure is,
\begin{multline}
H_{x} \equiv -\int_{\mathcal{D}_x} \rho_{x}(\vec{x})\, \textrm{log}\left(\rho_{x}(\vec{x})\right) d^D x\, \\
=\, H_{\mathcal{X}}-\int_{\mathcal{D}_x} \rho_{x}(\vec{x})\, \textrm{log}\left(l_P^{-D}\sqrt{\det(g_x(\vec{x}))}\right)\, d^D x\ , \label{hxha}
\end{multline}
which is not invariant under reparametrizations of the local coordinates,
\begin{equation}
H_{x} = H_{y}\, +\, \int_{\mathcal{D}_y} \rho_{y}(\vec{y})\, \textrm{log}\left(J_{\vec{x}\to\vec{y}}(\vec{y})\right) d^D y\label{hxt}
\end{equation}
The non-invariance of $H_x$ under reparametrizations is well-known in statistical inference, where the problem is circumvented by considering ratios of likelihoods \cite{pawitan2001all}. The ratio $\rho_{x}(\vec{x};\, \theta_1)/\rho_x(\vec{x};\, \theta_2)$ between the local probability density functions corresponding to two different states, $\theta_1$ and $\theta_2$, of the object measured by Alice is invariant under coordinate reparametrizations of the local chart, and therefore their relative entropy, given by the Kullback-Leibler divergence of the two probability densities, is also invariant. However, the non-invariance of $H_x$ is not an inherent property of information but a direct consequence of not properly accounting for the contribution of the geometry of $X$. Thus, in this work, we take a different approach and consider $H_{\mathcal{X}}$, defined in eq.~(\ref{hx}), which is invariant under coordinate reparametrizations and represents a measure of information content independent of the observer's reference frame.

\section{Microstate description of $H_\mathcal{X}$}\label{sec2}

We now consider the information gain associated with Alice's observations from the perspective of the classical microstates of the test particle she uses. These microstates span a $2D$-dimensional phase space given by the co-tangent bundle of $X$, $T^*X$. The exponential charts in $X$ can be naturally lifted to charts in $T^*X$. The co-tangent bundle is endowed with a natural symplectic structure, which in the local coordinates $\{x^\mu,p_\mu\}$ is given by the symplectic 2-form
\begin{equation*}
\omega = \frac{1}{h}\sum_\mu dx^\mu \wedge dp_\mu
\end{equation*}
where the position ($x^\mu$) and momentum ($p_\mu$) coordinates are conjugate variables and we have made $\omega$ dimensionless by using the Planck constant $h$. The volume measure in the phase space is given by the $2D$-form $d\,T^*X\equiv\frac{1}{D}\wedge^D \omega$, which reduces to the Lebesgue measure in the charts of $T^*X$. 

Assuming the information associated with Alice's observations has a microscopic realization, we expect the Gibbs entropy of the probe used by Alice to be related to $H_{\mathcal{X}}$. Classically, 
\begin{multline}
H_{\textrm{Gibbs}}\, \equiv \, -\int_{T^*X}\rho\, \textrm{log}(\rho)\, d\,T^*X \\
= -\frac{1}{h^D}\int_{\mathcal{D}_{x}\times\mathbb{R}^D} \rho_{x,p}(\vec x, \vec p)\, \textrm{log}(\rho_{x,p}(\vec x, \vec p)) \, d^Dx\, d^Dp\label{hxc}
\end{multline}
where the probability density function $\rho$ is dimensionless and describes the likelihood of finding the probe in the microstate $(\vec x, \vec p)$ at time $t_0$. Like $H_{\mathcal{X}}$, this expression is invariant under coordinate transformations. To see this, note that a reparametrization of the local coordinates induces a symplectic transformation in $T^*X$ that leaves invariant the measure $dT^*X$,
\begin{equation}
d^Dx = J_{\vec{x}\to\vec{y}}(\vec{y})\, d^Dy\ , \quad d^Dp = \left[J_{\vec{x}\to\vec{y}}(\vec{y})\right]^{-1} d^Dq \label{tpq}
\end{equation}
From the normalization condition $\int_{T^*X}\rho\, d\,T^*X=1$, it then follows that $\rho$, and consequently eq.~(\ref{hxc}), are also invariant.

To find the relation between $H_{\textrm{Gibbs}}$ and $H_{\mathcal{X}}$, we identify the probability density function $\rho_{x}(\vec{x})$, given in eq.~(\ref{phat}), with the marginal probability density function that results from integrating out the momentum degrees of freedom of $\rho_{x,p}(\vec{x},\vec{p})$ given by,
\begin{equation}
\rho_{x}(\vec{x})=\frac{1}{h^D}\int_{\mathbb{R}^D} \rho_{x,p}(\vec x, \vec p\,)\, d^Dp\label{marginal1}
\end{equation}
Under reparametrizations of the local coordinates, eq.~(\ref{tpq}) implies that the right-hand side of eq.~(\ref{marginal1}) transforms according to eq.~(\ref{tphat}). Hence, from this microscopic perspective, the Riemannian measure in the definition of $\rho_x(\vec{x})$ emerges from integrating out the momentum degrees of freedom, as the phase-space measure $dT^*X$ does not explicitly depend on the local metric $g_x$. This is consistent with the computation of the phase-space volume of a test particle near a Schwarzschild black hole \cite{padmanabhan1989phase}, where a similar observation was made.

Introducing the conditional probability density function in a chart of $T^*X$,
\begin{equation*}
\rho_{x,p}(\vec{p}\, |\, \vec{x})\equiv \frac{\rho_{x,p}(\vec x, \vec p)}{h^D\rho_{x}(\vec{x})}\ ,
\end{equation*}
we observe that $H_{\textrm{Gibbs}}$ can be decomposed as 
\begin{equation}
H_{\textrm{Gibbs}}= H_{\mathcal{X}} + H_{\mathcal{P}|\mathcal{X}}\label{gibbsdeco}
\end{equation}
with 
\begin{widetext}
\begin{equation*}
H_{\mathcal{P}|\mathcal{X}}\equiv -\frac{1}{h^D}\int_{\mathcal{D}_x\times \mathbb{R}^D}\rho_{x,p}(\vec{x},\vec{p})\,\textrm{log}\left[(l_P^{-1}h)^D\rho_{x,p}(\vec{p}\, | \,\vec{x})\,\sqrt{\det\left(g_x(\vec{x})\right)}\right] d^Dx\, d^Dp
\end{equation*}
\end{widetext}
Each term in this decomposition is invariant under reparametrizations of the local coordinates. This differs from the analogous decomposition of $H_{\textrm{Gibbs}}$ that results from using the Lebesgue measure \cite{hnizdo2010thermodynamic}, $H_{\textrm{Gibbs}} = H_x+H_{p|x}$, where the non-trivial transformation of $H_x$ under coordinate reparametrizations, eq.~(\ref{hxt}), is canceled with the non-trivial transformation of $H_{p|x}$.

A particular situation emerges when the position and momentum degrees of freedom are independent of each other. In that case,  $\rho$ factors out into marginal densities, $\rho_{x,p}(\vec{x},\vec{p}\,)=h^D\rho_x(\vec{x}\,)\rho_p(\vec{p}\,)$, with
\begin{equation}
\rho_p(\vec{p})=\frac{1}{h^D}\int_{\mathcal{D}_x}\rho_{x,p}(\vec{x},\vec{p})\, d^Dx\label{marginalp}
\end{equation}
and $\rho_x(\vec{x}\,)$ given in eq.~(\ref{marginal1}). The invariant conditional momentum entropy then becomes
\begin{multline}
H_{\mathcal{P}|\mathcal{X}}= H_p-\int_{\mathcal{D}_x}\rho_x(\vec{x}\,)\log\left[(l_P^{-1}h)^D\!\sqrt{\det\left(g_x(\vec{x})\right)} \right]d^Dx\\
\equiv H_{\mathcal{P}}\label{hhpp}
\end{multline}
where $H_p$ is the momentum differential entropy that Alice would conventionally define in her laboratory,
\begin{equation*}
H_p\equiv -\int_{\mathbb{R}^D}\rho_p(\vec{p}\,)\log\left(\rho_p(\vec{p}\,)\right)d^Dp\ .
\end{equation*}
Eq. (\ref{hhpp}) is the momentum counterpart of eq.~(\ref{hxha}), where the role of $l_P$ is now taken by the Planck momentum, $l_P^{-1}h$. In the specific case where $g$ is the Euclidean metric, $H_{\mathcal{P}}=H_p-D\log(l_P^{-1}h)$ and $H_{\mathcal{X}}=H_x-D\log(l_P)$.

Like $H_{\mathcal{P}|\mathcal{X}}$, the momentum entropy $H_\mathcal{P}$ is invariant under coordinate transformations. Since a coordinate transformation $x^{\mu}=x^{\mu}(\vec{y})$ needs to be accompanied by a reparametrization of the momentum coordinates, $q_{\mu}=\sum_\nu \frac{\partial x^\nu(\vec{y})}{\partial y^\mu} p_{\nu}$, for the new coordinates $y^\mu$ and momenta $q_\mu$ to be canonically conjugate variables, the probability density $\rho_p(\vec{p})$ transforms as
\begin{equation}
\rho_p(\vec{p})=E_{\vec{y}|\vec{q}}\left[J_{\vec{x}\to\vec{y}}(\vec{y})\right]\!(\vec{q})\, \rho_q(\vec{q})\label{trasp}
\end{equation}
where the conditional expectation is defined by
\begin{equation*}
E_{\vec{y}|\vec{q}}\left[f(\vec{y})\right](\vec{q})\equiv \frac{1}{h^D}\int_{\mathcal{D}_y} \frac{\rho_{y,q}(\vec{y},\vec{q})}{\rho_q(\vec{q})}f(\vec{y})\,d^Dy
\end{equation*}
Thus, under a reparametrization of the local coordinates, the transformation of the two integrals in eq.~(\ref{hhpp}) cancels between them, and $H_{\mathcal{P}}$ remains invariant.

\section{Quantum phase space}
\label{sec3}

We will now extend the classical invariant information framework introduced in Section \ref{sec2} to the quantum domain. The von Neumann entropy, $H^{\rm vN}=\textrm{Tr}\left[\pmb{\rho}\log(\pmb{\rho})\right]$, is widely regarded as the quantum analog of Gibbs entropy, where $\pmb{\rho}$ is the density matrix operator and the trace is evaluated over a complete orthonormal basis of the Hilbert space on which $\pmb{\rho}$ acts. However, $H^{\rm vN}$ does not account for the position and momentum degrees of freedom of the quantum state, solely depending on the statistical ensemble used to prepare the state and vanishing for pure states. Here, we adopt a different approach to extending eq.~(\ref{hxc}) into the quantum domain by considering the quantum analog of the classical phase space.

For simplicity, we consider the test particle used by Alice to be spinless and assume its mass ($m$), momentum, and couplings are small, such that quantum fluctuations of external fields and relativistic effects can be ignored. Under these assumptions, the one-particle approximation of quantum mechanics is valid, and the Hamiltonian operator in the Scr\"odinger picture is given by \cite{dewitt1952point, dewitt1957dynamical}
\begin{widetext}
\begin{equation*}
\mathcal{H}=
\frac{1}{2m}\sum_{\mu,\nu}\left(\left[\det\left(g(\vec{\mathbf{X}})\right)\right]^{-\frac14}\left(\vec{\mathbf{P}}-\vec{A}(\vec{\mathbf{X}})\right)_\mu (g(\vec{\mathbf{X}}))^{\mu\nu}\sqrt{\det\left(g(\vec{\mathbf{X}})\right)}\left(\vec{\mathbf{P}}-\vec{A}(\vec{\mathbf{X}})\right)_\nu \left[\det\left(g(\vec{\mathbf{X}})\right)\right]^{-\frac14}\right)
+V(\vec{\mathbf{X}})
\end{equation*}
\end{widetext}
Here, $\vec{A}(\vec{x})$ and $V(\vec{x})$ represent, respectively, the classical vector and scalar potentials induced by external gauge fields and non-kinetic gravitational interactions. The position ($\vec{\mathbf{X}}$) and momentum ($\vec{\mathbf{P}}$) operators adhere to the canonical commutation relations, $[\mathbf{X}^\mu,\mathbf{P}_{\nu}]=i\hbar\,\delta^{\mu}{}_{\nu}$ and $[\mathbf{X}^\mu,\mathbf{X}^\nu]=[\mathbf{P}_\mu,\mathbf{P}_\nu]=0$. 

The eigenvalues of $\vec{\mathbf{X}}$ span the entire Riemannian space $X$. We denote the eigenstate corresponding to the point with coordinates $\vec{x}$ in the chart $(U_x,\varphi_x)$ as $|\vec{x}\,\rangle_x$, normalized with respect to the Riemannian measure in $X$. We adopt a passive perspective on diffeomorphisms, where $|\vec{y}\,\rangle_y$, with $\{x^\mu\}\to\{y^\mu\}$ representing a reparametrization of the local coordinates, corresponds to the same physical state as $|\vec{x}\,\rangle_x$. Similarly, we denote the eigenstates of the canonical momentum operator by $|\vec{p}\,\rangle_{p}$, with momentum eigenvalue $\vec{p}$ in the chart of the cotangent bundle with coordinates $\{x^\mu, p_\mu\}$. Since the momentum and Hamiltonian operators typically do not commute, even if $V=\vec{A}=0$, the momentum eigenstates $|\vec{p}\,\rangle_{p}$ are generally not eigenstates of the free-particle Hamiltonian.

Under these conventions, the probability density function that describes the likelihood of finding the test particle at position $\vec{x}$ is given by
\begin{equation}
\hat{\rho}_x(\vec{x}\,)= {}_x\langle\vec{x}\,|\pmb{\rho}|\vec{x}\,\rangle_x=|\psi_x(\vec{x}\,)|^2\, , \label{qrhox}
\end{equation}
where $\pmb{\rho}\equiv |\psi\rangle\langle\psi|$ is the density matrix operator associated with $|\psi\rangle$, and $\psi_x(\vec{x})\equiv {}_x\langle\vec{x}\,|\psi\rangle$ is the position wavefunction representation of the state $|\psi\rangle$ in the chart $(U_x,\varphi_x)$ with respect to the dimensionless Riemannian measure in $X$. This function is the quantum analog of the classical probability density function introduced in eq.~(\ref{mua}). In particular, it is invariant under reparametrizations of the local coordinates, $\hat{\rho}_y(\vec{y})=\hat{\rho}_x(\vec{x}(\vec{y}))$. 

In the position wavefunction representation, the canonical position and momentum operators act as follows \cite{dewitt1952point}
\begin{align}
 (\hat{\mathbf{X}}_x)^\mu\psi_x(\vec{x}\,)&\equiv {}_x\langle \vec{x}\,|\mathbf{X}^\mu|\psi\rangle=x^\mu\, \psi_x(\vec{x})\, , \nonumber\\
 (\hat{\mathbf{P}}_x)_\mu\psi_x(\vec{x}\,)&\equiv {}_x\langle \vec{x}\,|\mathbf{P}_\mu|\psi\rangle\nonumber\\
 & =-i\hbar\left(\frac{\partial}{\partial x^\mu}+\frac14\frac{\partial\,\log\left[\det(g_x(\vec{x}))\right]}{\partial x^\mu}\right)\psi_x(\vec{x})\label{momentum}
\end{align}
The wavefunctions of the position and momentum eigenstates are therefore given by
\begin{align}
{}_x\langle \vec{x}\,'|\vec{x}\,\rangle_x&=l_P^D\,[\det(\vec{x})]^{-1/2}\, \delta^D(\vec{x}-\vec{x}\,')\, ,  \nonumber\\ 
{}_x\langle \vec{x}\,|\vec{p}\,\rangle_p &=\frac{(l_Ph^{-1})^{\frac{D}{2}}e^{\frac{i\vec{p}\cdot\vec{x}}{\hbar}}}{\left[\det\left(g_x(\vec{x}\,)\right)\right]^{\frac14}}\, ,\label{waves1}
\end{align}
with
\begin{align*}
(\hat{\mathbf{X}}_x)^\mu\, {}_x\langle \vec{x}\,'|\vec{x}\,\rangle_x &= x^\mu\, {}_x\langle \vec{x}\,'|\vec{x}\,\rangle_x\, ,  \nonumber\\ 
(\hat{\mathbf{P}}_x)_\mu\, {}_x\langle \vec{x}\,|\vec{p}\,\rangle_p &= p_\mu\, {}_x\langle \vec{x}\,|\vec{p}\,\rangle_p\, ,
\end{align*}
and ${}_p\langle \vec{p}\,'|\vec{p}\,\rangle_p = \delta^D(\vec{p}-\vec{p}\,')$. Using these expressions, we obtain the momentum wavefunction representation of $|\psi\rangle$ as
\begin{multline}
\psi_p(\vec{p}\,)\equiv {}_p\langle\vec{p}\,|\psi\rangle\\
=\frac{1}{l_P^D}\int_{\mathcal{D}_x} {}_p\langle\vec{p}\,|\vec{x}\,\rangle_x\, {}_x\langle\vec{x}\,|\psi\rangle \sqrt{\det(g_x(\vec{x}))}\, d^D x\\
=\frac{1}{(l_Ph)^{\frac{D}{2}}}\int_{\mathcal{D}_x} e^{\frac{-i\vec{p}\cdot\vec{x}}{\hbar}}\psi_x(\vec{x})\left[\det\left(g_x(\vec{x}\,)\right)\right]^{\frac14} d^Dx\, . \label{psip}
\end{multline}
The probability density function describing the likelihood of observing the test particle with momentum $\vec{p}$ is then given by $\rho_p(\vec{p}\,)= {}_p\langle\vec{p}\,|\pmb{\rho}|\vec{p}\,\rangle_p=|\psi_p(\vec{p}\,)|^2$.

To extend eq.~(\ref{hxc}) to the quantum domain, we must identify the quantum analog of the phase-space probability density function $\rho(\vec{x},\vec{p})$. A suitable approach is through the deformation quantization formulation of quantum mechanics \cite{zachos2002deformation}. In this formulation, observables are represented by c-number functions (classical kernels) in phase space, which is endowed with a non-commutative product (Moyal's $\star$-product). The equivalence of this framework to the standard wavefunction formulation of quantum mechanics is established by Weyl's correspondence, which maps Weyl-ordered operators in Hilbert space to classical kernels in phase space \cite{weyl1950theory, Groenewold:1946kp}. Under this correspondence, the density matrix operator $\pmb{\rho}$ is mapped to the Wigner phase-space quasiprobability density function, which represents the quantum counterpart of $\rho(\vec{x},\vec{p})$ \footnote{Note that the argument used here to introduce the quasiprobability density function $\rho(\vec{x},\vec{p})$ differs from the classical Radon-Nikodym-based reasoning employed in Section \ref{sec1}. If an argument more closely paralleling the classical one is possible in the quantum phase space, it would likely involve a non-commutative version of the Radon-Nikodym theorem.}. 

The Wigner function can be conveniently defined as the Fourier inverse of the expected value of the characteristic operator \cite{moyal1949quantum},
\begin{widetext}
\begin{multline}
\rho_{x,p}(\vec{x},\vec{p})= \left(\frac{\hbar}{2\pi}\right)^D\int_{\mathcal{D}_x\times\mathbf{R}^D}\langle \psi|e^{-\frac{i}{2}\vec{x}\,{}'\cdot\vec{\mathbf{P}}}e^{-i\vec{p}{}'\,\cdot\vec{\mathbf{X}}}e^{\frac{i}{2}\vec{x}{}'\,\cdot\vec{\mathbf{P}}}|\psi\rangle\, e^{-i(\vec{x}\,{}'\cdot\vec{p}+\vec{x}\,\cdot\vec{p}\,{}')}\,d^Dx'd^Dp'\\
=(l_P^{-1}\hbar)^D \sqrt{\det(g_x(\vec{x}))}\int_{\mathcal{D}_x} \langle\psi|e^{-\frac{i}{2}\vec{x}\,{}'\cdot\vec{\mathbf{P}}}|\vec{x}\,\rangle_x\,{}_x\langle\vec{x}\,|e^{\frac{i}{2}\vec{x}\,{}'\cdot\vec{\mathbf{P}}}|\psi\rangle\, e^{-i\vec{x}\,{}'\cdot\vec{p}}\,d^D\vec{x}\,{}'\ .\label{wigner1}
\end{multline}
\end{widetext}

To extend eq.~(\ref{hxc}) to the quantum domain, we need to formulate the Wigner function in arbitrary Riemannian spaces. While we could use eqs.~(\ref{waves1}) to write the matrix elements in this expression in terms of the position wavefunction $\psi_x(\vec{x}\,)$, we find it more convenient to follow a different but equivalent path. Recall from eq.~(\ref{marginal1}) that, at the classical level, integrating out the momentum degrees of freedom in $\rho(\vec{x},\vec{p}\,)$ results in a marginal probability density function with respect to the Lebesgue measure in $\mathcal{D}_x$ rather than the invariant probability density function $\hat\rho_x(\vec{x}\,)$. The operators $\hat{\mathbf{X}}_x$ and $\hat{\mathbf{P}}_x$, defined in eq.~(\ref{momentum}), are self-adjoint with respect to the inner product induced by the Riemannian measure, and give rise to the invariant function $\hat\rho_x(\vec{x}\,)$, eq.~(\ref{qrhox}). Following the same reasoning as in Section \ref{sec1}, we can define new position eigenstates $|\vec{x}\,\rrbracket_x\equiv l_P^{-\frac{D}{2}}\left[\det\left(g_x(\vec{x}\,)\right)\right]^{\frac14}|\vec{x}\,\rangle_x$ normalized with respect to the inner product induced by the Lebesgue measure in $\mathcal{D}_x$. In this basis, the position wavefunction representation of $|\psi\rangle$ is $\psi_x^\lambda(\vec{x}\,)\equiv {}_x\llbracket\vec{x}\,|\psi\rangle_x=l_P^{-\frac{D}{2}}\left[\det\left(g_x(\vec{x}\,)\right)\right]^{\frac14}\psi(\vec{x}\,)$. We refer to these wavefunctions as $\lambda$-wavefunctions. The probability density function describing the likelihood of finding the test particle at position $\vec{x}$ with respect to the Lebesgue measure, introduced in eq.~(\ref{phat}), is therefore given at the quantum level by
\begin{equation*}
\rho_x(\vec{x}\,)= {}_x\llbracket\vec{x}\,|\pmb{\rho}|\vec{x}\,\rrbracket_x=|\psi^\lambda_x(\vec{x}\,)|^2\, . 
\end{equation*}

In the $\lambda$-wavefunction representation, the canonical position and momentum operators act as
\begin{align*}
 (\mathbf{X}_x)^\mu\psi^\lambda_x(\vec{x}\,)&\equiv {}_x\llbracket \vec{x}\,|\mathbf{X}^\mu|\psi\rangle=x^\mu\, \psi^\lambda_x(\vec{x})\, , \nonumber\\
 (\mathbf{P}_x)_\mu\psi^\lambda_x(\vec{x}\,)&\equiv {}_x\llbracket \vec{x}\,|\mathbf{P}_\mu|\psi\rangle=-i\hbar\frac{\partial\psi^\lambda_x(\vec{x})}{\partial x^\mu}
\end{align*}
and the position and momentum $\lambda$-wavefunctions are
\begin{align*}
{}_x\llbracket \vec{x}\,'|\vec{x}\,\rrbracket_x&=\delta^D(\vec{x}-\vec{x}\,')&(\mathbf{X}_x)^\mu\, {}_x\llbracket \vec{x}\,'|\vec{x}\,\rrbracket_x &= x^\mu\, {}_x\llbracket \vec{x}\,'|\vec{x}\,\rrbracket_x\, ,  \\ 
{}_x\llbracket \vec{x}\,|\vec{p}\,\rangle_p &=\frac{1}{h^\frac{D}{2}}e^{\frac{i\vec{p}\cdot\vec{x}}{\hbar}}& (\mathbf{P}_x)_\mu\, {}_x\llbracket \vec{x}\,|\vec{p}\,\rangle_p &= p_\mu\, {}_x\llbracket \vec{x}\,|\vec{p}\,\rangle_p\, ,
\end{align*}
In particular, $\mathbf{X}_x$ and $\mathbf{P}_x$ are self-adjoint with respect to the inner product induced by the Lebesgue measure in $\mathcal{D}_x$, and the momentum wavefunction, given in eq.~(\ref{psip}), can be equivalently expressed in terms of the $\lambda$-wavefunction as
\begin{multline}
\psi_p(\vec{p}\,)= 
\int_{\mathcal{D}_x} {}_p\langle\vec{p}\,|\vec{x}\,\rrbracket_x\, {}_x\llbracket\vec{x}\,|\psi\rangle\,  d^D x\\
=\frac{1}{h^{\frac{D}{2}}}\int_{\mathcal{D}_x} e^{-i\frac{\vec{p}\cdot\vec{x}}{\hbar}}\psi_x^\lambda(\vec{x})\, d^Dx 
\label{lambdamom}
\end{multline}

We can utilize the $\lambda$-wavefunction representation to express the matrix elements appearing in eq.~(\ref{wigner1}) as
\begin{equation*}
{}_x\langle\vec{x}\,|e^{i\vec{\tau}\cdot\vec{\mathbf{P}}}|\psi\rangle=\frac{l_P^{\frac{D}{2}}\,{}_x\llbracket\vec{x}\,|e^{i\vec{\tau}\cdot\vec{\mathbf{P}}}|\psi\rangle}{\left[\det\left(g_x(\vec{x}\,)\right)\right]^{\frac{1}{4}}}=\frac{l_P^{\frac{D}{2}}\psi_x^\lambda(\vec{x}+\hbar\vec{\tau})}{\left[\det\left(g_x(\vec{x}\,)\right)\right]^{\frac{1}{4}}}\, ,
\end{equation*}
which leads us to the following compact expression for the Wigner quasiprobability density function in arbitrary Riemannian spaces,
\begin{multline}
\rho_{x,p}(\vec{x},\vec{p})=\\
\!\int_{\mathcal{D}_x} \left[\psi_x^\lambda\right]^*\!\left(\vec{x}-\frac{\vec{x}\,{}'}{2}\right)\psi^\lambda_x\left(\vec{x}+\frac{\vec{x}\,{}'}{2}\right)e^{-i\frac{\vec{p}\cdot\vec{x}{}'}{\hbar}}\,d^D\vec{x}\,{}'\label{quantrho}
\end{multline}
Thus, when expressed in terms of $\lambda$-wavefunctions rather than conventional wavefunctions normalized with respect to the Riemannian measure, the expression for the Wigner function in arbitrary Riemannian spaces has the same form as in flat space \cite{zachos2002deformation}. This expression is mathematically equivalent to that introduced in \cite{gneiting2013quantum} by using translations to construct Stratonovich-Weyl operator kernels.

As expected, integrating over the position or momentum degrees of freedom in eq.~(\ref{quantrho}) yields the marginal probability density functions introduced in eqs.~(\ref{marginal1}) and (\ref{marginalp}), with $\rho_p(\vec{p}\,)=|\psi_p(\vec{p}\,)|^2$ and $\rho_x(\vec{x}\,)=l_P^{-D}\sqrt{\det\left(g_x(\vec{x}\,)\right)}\,|\psi_x(\vec{x}\,)|^2$. Under reparametrizations of the local coordinates, this expression for the Wigner function remains invariant, $\rho_{x,p}(\vec{x},\vec{p}\,)=\rho_{y,q}(\vec{y}(\vec{x}),\vec{q}(\vec{p}))$, while the marginal density functions $\rho_x(\vec{x}\,)$ and $\rho_p(\vec{p}\,)$ transform according to eqs.~(\ref{tphat}) and (\ref{trasp}), respectively. More generally, the probability density function with respect to the Lebesgue measure for a physical observable represented by the classical kernel $g(\vec{x},\vec{p}\,)$ is given by
\begin{multline}
\rho_g(\xi)=\\
\frac{1}{2\pi h^D}\int_{\mathcal{D}_x\times\mathbb{R}^{D+1}}e_{\star}^{i t g(\vec{x},\vec{p}\,)}\, e^{-it\xi}\,\rho_{x,p}(\vec{x},\vec{p}\,)\, d^Dx\,d^Dp\,dt
\label{rhogg}
\end{multline}
where the $\star$-exponential is defined as $e_\star^A\equiv 1+A+\frac{1}{2!}A\star A+\frac{1}{3!}A\star A\star A+\ldots$ and $\star\equiv e^{\frac{i\hbar}{2}(\cev{\partial}_x\cdot\vec{\partial}_p-\cev{\partial}_p\cdot\vec{\partial}_x)}$ \cite{zachos2002deformation}.

The expression for the Wigner function in arbitrary Riemannian spaces presented in eq.~(\ref{quantrho}) differs from the one proposed in \cite{alonso2002wigner, alonso2003wigner} for spaces of constant curvature. In arbitrary Riemannian spaces, momentum is generally not conserved, and various physically distinct operators can be adopted that converge to the ordinary momentum in the limit of vanishing curvature. For maximally symmetric spaces, as considered in \cite{alonso2002wigner, alonso2003wigner}, the eigenstates of the Laplace-Beltrami operator are also eigenstates of the Killing vector fields. These Killing vector fields can be used to define a conserved momentum operator in such spaces. However, this momentum operator does not satisfy canonical conjugation relations with the position operator. As a result, the Wigner quasiprobability density function constructed from this conserved momentum operator, while useful in the context of maximally symmetric spaces, is not invariant under canonical transformations.

\section{Quantum phase-space entropy}\label{sec4}

We still must address an additional difficulty in extending the invariant framework of Section \ref{sec2} to the quantum domain. Unlike wavefunctions, the Wigner quasiprobability density function depends on both position and momentum degrees of freedom, making the symplectic symmetry of the theory manifest. However, this manifest invariance comes at a cost: by doubling the degrees of freedom, the Wigner function can take negative values.  

Negative quasiprobabilities in quantum mechanics are necessary for interpreting the contributions of unobservable virtual quantum states to physical observables in terms of classical phase space \cite{feynman1984negative}. To maintain equivalence between the wavefunction and phase-space descriptions, an additional ``on-shell'' constraint must be imposed. For any physical observable, the probability density function describing the possible outcomes of a measurement (c.f. eq.~(\ref{rhogg})) must adhere to the axioms of Kolmogorov's probability theory. In the case of the Wigner function, this is ensured by their functional form, eq.~(\ref{wigner1}) \cite{zachos2002deformation}. 

Having identified $\rho(\vec{x},\vec{p}\,)$ with the Wigner quasiproability density function at the quantum level, the expression for Gibbs entropy introduced in eq.~(\ref{hxc}) can be only applied to quantum states with a non-negative quasiprobability density, known as \emph{Wigner-positive states} \cite{van2021quantum}, as the entropy becomes complex and multi-valued for $\rho(\vec{x},\vec{p}\,)<0$. However, if there is an invariant notion of quantum information for Riemannian spaces, a natural generalization of eq.~(\ref{hxc}) to quasiprobability density functions should exist, accounting for the contribution of intermediate virtual quantum states to entropy.

Shannon derived his renowned formula for the entropy of a discrete probability distribution by ensuring that the entropy function is continuous with respect to the probabilities, increases monotonically with the number of equally probable states, and can be represented as a weighted sum of individual entropies when a decision process can be decomposed into successive choices (the chain rule of conditional entropy) \cite{shannon1948mathematical}. To extend Shannon's entropy to quasiprobability distributions, we need to find an expression that continues to satisfy these conditions and reduces to Shannon's expression when all quasiprobabilities are in the unit interval $[0, 1]$.

\begin{figure}[!ht]
  \centering
  \includegraphics[width=70mm]{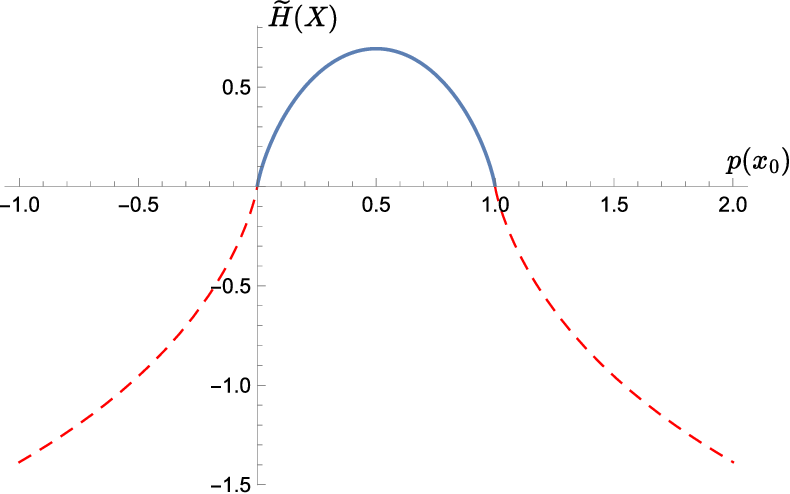}
  \caption{Entropy of a two-state quasirandom variable as a function of the quasiprobability of one of the states. States with negative quasiprobability have negative entropy (highlighted in dashed red).}\label{fig2}
\end{figure}

Consider a quasirandom variable $X$ with states $\{x_i\}$ and quasiprobabilities $p(x_i)$. The function $\widetilde{H}(X)\equiv -\sum_ip(x_i)\log|p(x_i)|$ reduces to Shannon's entropy when $p(x_i)\in [0,1]$. This function is continuous over $p(x_i)\in \mathbb{R}$ and smooth over $p(x_i)\in \mathbb{R}\backslash\{0,1\}$. Furthermore, if $X$ and $Y$ are two quasirandom variables with joint quasiprobabilities $p(x_i,y_j)=p(y_j)\,p(x_i|y_j)$, $\widetilde{H}(X,Y)$ satisfies the chain rule of conditional entropy, $\widetilde{H}(X,Y)=\widetilde{H}(Y)+\sum_j p(y_j)\,\widetilde{H}(X|Y\!\!\!=\!\!\!y_j)$. In particular, this condition is not satisfied by other analytic continuations, such as $-\sum_i|p(x_i)|\log|p(x_i)|$. We thus refer to $\widetilde{H}(X)$ as the entropy of $X$. Fig. \ref{fig2} illustrates the entropy of a two-state quasirandom variable as a function of the quasiprobability of one of the states.

Based on the above, we define the phase-space entropy of a quantum state $|\psi\rangle$ as
\begin{equation}
H_{\mathcal{X},\mathcal{P}}\, \equiv \, -\frac{1}{h^D}\int_{\mathcal{D}_{x}\times\mathbb{R}^D} \rho_{x,p}(\vec x, \vec p)\, \textrm{log}\,|\rho_{x,p}(\vec x, \vec p)| \, d^Dx\, d^Dp \label{psentropy}
\end{equation}
with $\rho_{x,p}(\vec x, \vec p)$ given by eq.~(\ref{quantrho}). In particular, $H_{\mathcal{X},\mathcal{P}}$ is invariant under symplectic transformations of the phase space and can be decomposed as in eq.~(\ref{gibbsdeco}). This expression can also be interpreted as the real part of the complex-valued entropy of the Wigner function, which has been studied in \cite{laguna2010shannon,salazar2023phase,cerf2024complex}. 

The phase-space entropy defined in eq.~(\ref{psentropy}) has a microscopic interpretation, as it does not involve any coarse-graining of the phase space, and it reduces to Gibbs entropy for classical states. When the system is prepared in a mixed state with density matrix operator $\pmb{\rho}^{\rm mixed}=\sum_a p_a |\psi_a\rangle\langle\psi_a|$, with $\sum_a p_a=1$ and $p_a\in[0,1]$, we can use the chain rule to compute the total entropy of the system, obtaining
\begin{equation*}
H^{\rm total}\, = \, H^{\textrm{vN}} + \sum_a p_a H_{\mathcal{X},\mathcal{P}}^{(a)}\ ,
\end{equation*}
where $H^{\textrm{vN}}=-\sum_a p_a\log(p_a)$ is the von Neumann entropy of $\pmb{\rho}^{\rm mixed}$ and $H_{\mathcal{X},\mathcal{P}}^{(a)}$ the phase-space entropy of $|\psi_a\rangle$. In particular, when the pure states $|\psi_a\rangle$ all have the same phase-space entropy, the total entropy of the system is given by von Neumann's entropy up to an additive constant that is independent of the state probabilities. 

Since the density matrix operator $\pmb{\rho}^{\rm mixed}$ results from integrating out the degrees of freedom of the environment in the system comprising the observed state along with its interacting environment \cite{zurek2003}, some of the information in the ensemble defined by the probabilities $\{p_a\}$ may be partially contained in the microscopic (position and momentum) degrees of freedom of the mixed state. This redundancy can be quantified by the mutual information between the ensemble and the microscopic degrees of freedom of the mixed state,
\begin{equation*}
I=H_{\mathcal{X},\mathcal{P}}-\sum_a p_a H_{\mathcal{X},\mathcal{P}}^{(a)}\ .
\end{equation*}
where $H_{\mathcal{X},\mathcal{P}}$ is the phase-space entropy of the mixed state.

\section{Quantum phase-space entropy of the harmonic oscillator}
\label{sec5}

We now demonstrate the computation of the quantum phase-space entropy using a non-relativistic harmonic oscillator in 2-dimensional spacetime. We consider two different spacetime geometries: Minkowski and anti-de Sitter ($AdS_2$). 

\subsection{Flat space}
\label{minks}

\begin{figure}[!ht]
  \centering
  \includegraphics[width=70mm]{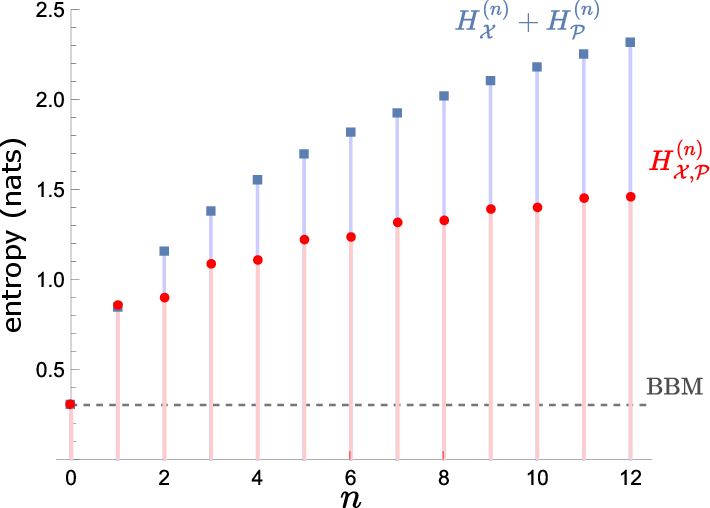}
  \caption{Phase-space entropy (red circles) and the sum of the invariant momentum and position entropies (blue squares) for the quantum harmonic oscillator in flat space as a function of the energy level. The dashed line indicates the BBM lower bound to $H_{x}+H_{p}$.}\label{osci}
\end{figure}

Consider a spinless test particle of mass $m$ influenced by a quadratic potential, with a coupling constant $\kappa$, in 2-dimensional Minkowski spacetime. In the position wavefunction representation, the Hamiltonian operator of the particle is
\begin{equation*}
\hat{\mathcal{H}}_x=\frac{(\hat{\mathbf{P}}_x)^2}{2m}+\kappa\frac{x^2}{2}
\end{equation*}
with $\hat{\mathbf{P}}_x=-i\hbar\frac{\partial}{\partial x}$. The eigenstates of this operator are
\begin{equation*}
\psi^{(n)}_x(x)=\sqrt{\frac{l_P}{2^n n!}}\left(\frac{\sqrt{\kappa m}}{\pi\hbar}\right)^{1/4}\!H_n\left(\sqrt{\frac{\sqrt{\kappa m}}{\hbar}}x\right)\, e^{-\frac{\sqrt{\kappa m}}{2\hbar}x^2}
\end{equation*}
with $H_n(x)$ the Hermite polynomial or order $n$, corresponding to the energy level $E^{(n)}=\hbar\sqrt{\frac{\kappa}{m}}(n+\frac12)$. In flat space, $\lambda$-wavefunctions and ordinary wavefunctions coincide up to a factor $l_P^{1/2}$, and eq.~(\ref{quantrho}) reduces to the standard Wigner function of the quantum harmonic oscillator in flat space \cite{zachos2002deformation},
\begin{multline*}
\rho_{x,p}^{(n)}(x,p)=\\
2(-1)^n L_n\left(\frac{2\sqrt{\kappa m}}{\hbar}x^2+\frac{2}{\hbar\sqrt{\kappa m}}p^2\right) e^{-\frac{\sqrt{\kappa m}}{\hbar}x^2-\frac{p^2}{\hbar\sqrt{\kappa m}}}
\end{multline*}
with $L_n(x)$ the Laguerre polynomial of order $n$. Inserting this expression into eq.~(\ref{psentropy}) and using the identity eq.~(\ref{int1}) derived in the Appendix, we obtain the following expression for the phase-space entropy of the energy eigenstates,
\begin{widetext}
\begin{equation*}
H^{(n)}_{\mathcal{X},\mathcal{P}}=1+2n-\log(2)+\sum_{q=0}^n\sum_{\lambda\in\textrm{Roots}(L_n)}\sum_{l=0}^q\left(\begin{matrix}n\\ q\end{matrix}\right)\left[\frac{(-1)^{q+n}\,\lambda^l\,2^{q-l}}{l!}e^{-\frac{\lambda}{2}}Ei\left(\frac{\lambda}{2}\right)-\sum_{k=1}^l\sum_{s=0}^{k-1}\frac{(-1)^{s+q+n}\,\lambda^{l-k+s}\,2^{q-l-s+k}}{s!\,(l-k)!\,k}\right]
\end{equation*}
\end{widetext}
where $Ei(x)$ is the exponential integral function. Remarkably, all dimensionful quantities, including the Planck constant, the Plack length, the mass, and the oscillator coupling constant, have canceled in this expression. In particular, the phase-space entropies of the ground and first excited energy states are,
\begin{align}
H^{(0)}_{\mathcal{X},\mathcal{P}}&=1-\log(2)\simeq 0.307 \ \textrm{nats}\label{equat}\\
H^{(1)}_{\mathcal{X},\mathcal{P}}&=1-\log(2)+\frac{2}{\sqrt{e}}Ei\left(\frac{1}{2}\right)\simeq 0.858 \ \textrm{nats}\nonumber
\end{align}
We have represented in Fig. \ref{osci} the value of $H^{(n)}_{\mathcal{X},\mathcal{P}}$ for the first energy levels and compared it to the sum of the invariant momentum and position entropies, $H^{(n)}_{\mathcal{X}}+H^{(n)}_{\mathcal{P}}$ (c.f. eqs.~(\ref{hx}) and (\ref{hhpp})). For the ground state, the momentum and position degrees of freedom are independent and $H^{(0)}_{\mathcal{X},\mathcal{P}}=H^{(0)}_{\mathcal{X}}+H^{(0)}_{\mathcal{P}}$. However, for higher energy levels, $H^{(n)}_{\mathcal{X}}+H^{(n)}_{\mathcal{P}}$ grows faster than $H^{(n)}_{\mathcal{X},\mathcal{P}}$, indicating that the amount of physical information contained in the system is substantially reduced by the mutual information between the position and momentum degrees of freedom.

\subsection{Hyperbolic space}

We now consider the case where the spacetime geometry is $AdS_2$. The spacetime metric tensor in a global coordinate system is
\begin{equation*}
ds^2 = -c^2[\gamma(x)]^{-1}dt^2+\gamma(x) dx^2
\end{equation*}
where $\gamma(x)=\frac{R^2}{R^2+x^2}$, with $R$ the radius of curvature, and $c$ is the speed of light. The metric in a constant-time slice $X$ is thus $ds_X^2=\gamma(x) dx^2$. In particular, the spacelike component of the conformal boundary of $AdS_2$ consists of the 0-sphere generated by the two points $x=\pm\infty$.

In the position wavefunction representation, the Hamiltonian operator of the test particle is now
\begin{equation*}
\hat{\mathcal{H}}_x=\frac{(\hat{\mathbf{P}}_x)^2}{2m}+\kappa\gamma(x)\frac{x^2}{2}
\end{equation*}
with $\hat{\mathbf{P}}_x=-i\hbar [\gamma(x)]^{-\frac12}\frac{\partial}{\partial x}$. The eigenfunctions of this Hamiltonian have been computed in \cite{carinena2004one, carinena2007quantum}. Here, we focus on the ground state,
$\psi^{(0)}_x(x) = (l_P\,\mathcal{N})^{1/2}\,[\gamma(x)]^{1/2\Lambda}$, 
with energy $E^{(0)}=\hbar^2/(2mR^2\Lambda)$, where $\mathcal{N}\equiv \Gamma\left(\frac{1}{\Lambda}+\frac{1}{2}\right)/\left(\pi^{\frac12} R\, \Gamma\left(\frac{1}{\Lambda}\right)\right)$ and $\Lambda\equiv 2\hbar/(-\hbar+\sqrt{\hbar^2+4m\kappa R^4})$. The corresponding  $\lambda$-wavefunction is, 
\begin{equation}
\psi^{(0)\,\lambda}_x(x) = \mathcal{N}^{\frac{1}{2}}\,[\gamma(x)]^{\frac{1}{2\Lambda}+\frac{1}{4}}\label{wave0}
\end{equation}

This wavefunction is multi-valued unless $1/\Lambda$ takes the form $2j-\frac{1}{2}$ with $j\in \mathbb{N}^+$. This requirement leads to a constraint relating the mass and the coupling constant of the test particle to the curvature radius of $AdS_2$,
\begin{equation}
m\kappa=\frac{4\hbar^2}{R^4}\left(j+\frac{1}{4}\right)\left(j-\frac{1}{4}\right) \label{bps}
\end{equation}
The same condition ensures that the $\lambda$-wavefunctions of all energy eigenstates are single-valued. 

\begin{figure*}[!ht]
  \centering
  \includegraphics[width=112mm]{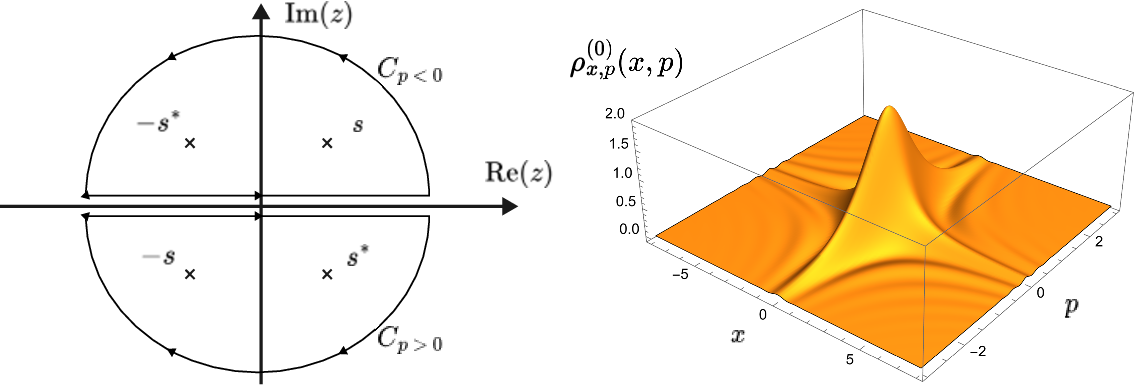}
  \caption{Left: Integration contour around the four poles of order $j$ of the Wigner quasiprobability density function $\rho^{(0)}_{x,p}(x,p)$ (eq.~(\ref{pole})) for $p>0$ ($C_{p\,>\,0}$) and $p<0$ ($C_{p\,<\,0}$). The radius of the semicircular part of the contour is taken to be $\infty$. Right: Plot of the quasiprobability density function $\rho^{(0)}_{x,p}(x,p)$ as a function of $x$ and $p$ for $j=R=\hbar=1$.}\label{fig1}
\end{figure*}

Inserting eq.~(\ref{wave0}) into eq.~(\ref{quantrho}), we obtain the Wigner quasiprobability density function for the ground energy state of the harmonic oscillator in $AdS_2$,
\begin{multline*}
\rho^{(0)}_{x,p}(x,p)\, =\\
\mathcal{N}R^{2j}\int_{-\infty}^{\infty}\frac{e^{-i\frac{yp}{\hbar}}dy}{\left[R^2+\left(x-\frac{y}{2}\right)^2\right]^{j}\left[R^2+\left(x+\frac{y}{2}\right)^2\right]^{j}}
\end{multline*}
Eq.~(\ref{bps}) ensures that the integrand can be analytically continued to a meromorphic function in the complex plane so that the integral can be performed in terms of residues,
\begin{multline}
\rho^{(0)}_{x,p}(x,p)=\\
(4R)^{2j}\mathcal{N}\int_C\frac{e^{-i\frac{zp}{\hbar}}dz}{(z-s)^j\, (z+s)^j\, (z-s^*)^j\, (z+s^*)^j}\label{pole}
\end{multline}
where $s=2(x+iR)$ and the integration contour $C$ is depicted in Fig.~\ref{fig1}. 

The integral is carried out in Appendix \ref{countour}. For the particular case of $j=1$,
\begin{multline}
\rho^{(0)}_{x,p}(x,p)= \\
2R^2\frac{e^{-\frac{2|p|R}{\hbar}}}{R^2+x^2}\left[\cos\left(\frac{2|p|x}{\hbar}\right)+\frac{R}{x}\sin\left(\frac{2|p|x}{\hbar}\right)\right]
\label{rhoads2}
\end{multline}
In particular, the quasiprobability density function of the ground state exhibits negative values in certain regions of the phase space and does not factor into independent momentum and position components, in contrast to what occurs for the ground state of the harmonic oscillator in flat space. We have plotted this quasiprobability density function for $R=\hbar=1$ in Fig.~\ref{fig1}.

As expected, integrating over the momentum or the position degrees of freedom produces the correct marginal probability density functions,
\begin{multline*}
\rho_x^{(0)}(x)\, =\, \frac{1}{h}\int_{-\infty}^\infty \rho^{(0)}_{x,p}(x,p)\,dp \\ 
=\,  \frac{1}{l_P}|\psi_x^{(0)}(x)|^2\sqrt{\gamma(x)}\, =\, \frac{2}{\pi}\frac{R^3}{(R^2+x^2)^2}\, 
\end{multline*}
and
\begin{multline*}
\rho_p^{(0)}(p)\, =\, \frac{1}{h}\int_{-\infty}^\infty \rho^{(0)}_{x,p}(x,p)\,dx \\ 
=\,\frac{1}{l_Ph}\left|\int_{-\infty}^{\infty}e^{-\frac{ipx}{\hbar}}[\gamma(x)]^{\frac{1}{4}}\psi_x^{(0)}(x)dx\right|^2 \, =\,\frac{R}{\hbar}e^{-\frac{2|p|R}{\hbar}}\ .
\end{multline*}

To compute the phase-space entropy, we insert eq.~(\ref{rhoads2}) into eq.~(\ref{psentropy}), leading to \footnote{In this case, we could only perform the integral semi-analytically, using the tanh-sinh quadrature method \cite{takahasi1974double} to integrate the following term,
\begin{multline*}
\int_0^{\infty}\int_0^{\infty}\frac{e^{-p}}{1+x^2}\left[\cos(px)+\frac{1}{x}\sin(px)\right]\cdot\\
\log\left(\left[\cos(px)+\frac{1}{x}\sin(px)\right]^2\right)\,dx\,dp \\
= \frac{\pi}{2}\quad \textrm{to within }<10^{-9}
\end{multline*}}
\begin{align*}
& &H^{(0)}_{\mathcal{X},\mathcal{P}}=\log(2)-\frac12\simeq 0.193\ \textrm{nats}  & \qquad \qquad (j=1)\, ,
\end{align*}
which again is independent of all dimensionful quantities. We thus observe that the information gain is lower in $AdS_2$ than in flat spacetime. This is largely due to an increment in the mutual information between the position and momentum degrees of freedom as a consequence of the spatial curvature. Specifically, for $j=1$,
\begin{equation}
H^{(0)}_{\mathcal{X},\mathcal{P}}=H^{(0)}_{\mathcal{X}}+H^{(0)}_{\mathcal{P}}+\frac{1}{2}\int_{-\infty}^{\infty} \rho_x^{(0)}(x)\,\log\left(\gamma(x)\right)\,dx
\label{conj1}
\end{equation}
with $H^{(0)}_{\mathcal{X}}=-3/2+\log(2)+\log(2\pi R/l_P)$ and $H^{(0)}_{\mathcal{P}}=1/2+\log(2)-\log(2\pi R/l_P)$. Thus, while $H^{(0)}_{\mathcal{X},\mathcal{P}}$ is independent of the curvature radius of $AdS_2$, $H^{(0)}_{\mathcal{X}}$ and $H^{(0)}_{\mathcal{P}}$ are not, allowing information to be exchanged between the position and momentum degrees of freedom through variations in the spatial curvature. 

\begin{figure}[!ht]
  \centering
  \includegraphics[width=55mm]{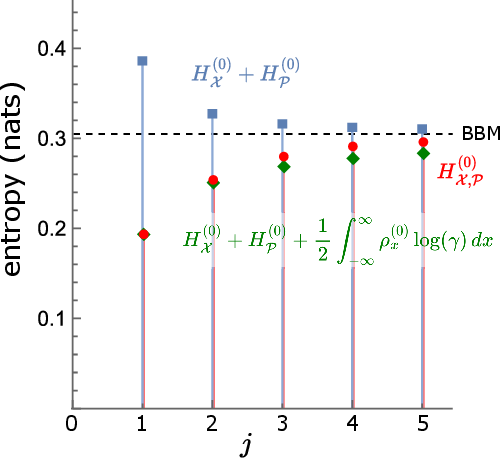}
  \caption{Phase-space entropy (red circles) and the sum of the invariant momentum and position entropies (blue squares) for the ground state of the quantum harmonic oscillator in $AdS_2$ as a function of $j$. The dashed line indicates the BBM lower bound to $H_{x}+H_{p}$ in flat space. The conjectured lower bound from eq.~(\ref{conjfin}) is also depicted (green diamonds).}\label{ads2fig}
\end{figure}

In Fig.~\ref{ads2fig}, we present the phase-space entropy alongside the combined invariant momentum and position entropies computed by numerical integration for $j>1$. As anticipated, for $j\gg 1$, both $H^{(0)}_{\mathcal{X},\mathcal{P}}$ and  $H^{(0)}_{\mathcal{X}}+H^{(0)}_{\mathcal{P}}$ converge towards the phase-space entropy of the ground state of the harmonic oscillator in flat space, eq.~(\ref{equat}).

\section{Quantum bounds on phase-space entropy}
\label{secbounds}

In 1957, Everett III \cite{dewitt2025many} and Hirschman \cite{hirschman1957note} independently conjectured a lower bound for the sum of position and momentum information entropies in flat space, expressed as $H_x+H_p\geq D(1+\log(\hbar\pi))$. This inequality implies Heisenberg's uncertainty relation and is saturated in cases where Heisenberg's inequality is not, providing a tighter lower bound to the product of position and momentum uncertainties. Bialynicki-Birula and Mycielski (BBM) \cite{bialynicki1975uncertainty} proved the inequality two decades later using Beckner's sharp Hausdorff-Young inequality for the Lebesgue norm of the Fourier transform \cite{beckner1975inequalities}. This result is now widely known as the BBM inequality. 

In a Riemannian space, the momentum wavefunction is related to the $\lambda$-wavefunction through an ordinary Fourier transformation, as noted in eq.~(\ref{lambdamom}). Thus, by applying the procedure of \cite{bialynicki1975uncertainty} to $\lambda$-wavefunctions, we can derive a generalized form of the BBM inequality for arbitrary Riemannian spaces,
\begin{equation}
H_{\mathcal{X}}+H_{\mathcal{P}}\geq D(1-\log(2))\label{bbmgen}
\end{equation}
where the invariant position and momentum entropies, $H_{\mathcal{X}}$ and $H_{\mathcal{P}}$, are defined in eqs.~(\ref{hx}) and (\ref{hhpp}), respectively. As expected, the ground state of the quantum harmonic oscillator in $AdS_2$ satisfies this inequality, as illustrated in Fig.~\ref{ads2fig}.

A tighter inequality than the BBM inequality has recently been conjectured for Wigner-positive states in flat space \cite{hertz2017entropy,van2021quantum}. Expressed in terms of invariant entropies, this inequality takes the form
\begin{equation}
H_\mathcal{X}+H_\mathcal{P}\geq H_{\mathcal{X},\mathcal{P}}\geq D(1-\log(2))
\label{conj}
\end{equation}
where the first inequality arises from Jensen's inequality, and the second is conjectured, with pure Gaussian states saturating it. Both inequalities can be violated by Wigner-negative states, that is, states with negative quasiprobability density in regions of the phase space. For example, the first excited state of the quantum harmonic oscillator in flat space, studied in Section \ref{minks}, violates the first inequality with
\begin{multline*}
H_{\mathcal{X}}^{(1)}+H_{\mathcal{P}}^{(1)}-H_{\mathcal{X},\mathcal{P}}^{(1)}=\\
-2+\log(4)-\frac{2}{\sqrt{e}}Ei\left(\frac12\right)+2\gamma
\simeq-0.01\ \textrm{nats}\ ,
\end{multline*}
where $\gamma$ is the Euler-Mascheroni constant. In general, violations of the first inequality in eq.~(\ref{conj}) are expected to be small, as the Wigner quasiprobability density must remain positive across most of the phase space to ensure it integrates to $h$. Likewise, numerical analyses have shown that Wigner-negative states in flat space can also violate the second inequality in eq.~(\ref{conj}) \cite{cerf2024complex}.

Extending eq.~(\ref{conj}) to arbitrary Riemannian spaces requires identifying the curved-space counterpart of Wigner-positive states. This is a non-trivial task since the spatial curvature typically induces a negative quasiprobability density in regions of the phase space, as we have seen for the ground state of the harmonic oscillator in $AdS_2$, where the second inequality in eq.~(\ref{conj}) is strongly violated (c.f. Fig.~\ref{ads2fig}). However, for the ground state of the harmonic oscillator and related coherent and squeezed states in arbitrary Riemannian spaces, we expect the additional contributions to the mutual information between the position and momentum degrees of freedom to arise solely from the metric. For such states, based on eq.~(\ref{conj1}), it appears natural to conjecture the following lower bound to $H_{\mathcal{X},\mathcal{P}}$,
\begin{equation}
H_{\mathcal{X},\mathcal{P}}\geq H_{\mathcal{X}}+H_{\mathcal{P}}+\int_{\mathcal{D}_x} \rho_x(\vec{x})\,\log\left(\sqrt{\det\left(g_x(\vec{x})\right)}\right)\,d^Dx \label{conjfin}
\end{equation}

Our numerical studies, summarized in Fig.~\ref{ads2fig}, confirm that the ground state of the harmonic oscillator in $AdS_2$ satisfies this inequality for arbitrary values of $j$.

\section{Discussion}
\label{discus}

We have investigated the consequences of requiring quantum information to be independent of the observer's reference frame. In the resulting formulation, quantum information possesses an absolute meaning inherently tied to the space (or phase space) it occupies. While our analysis has focused on stationary states, the unitary evolution of the quasiprobability density function ensures that the phase-space entropy of a closed system is conserved over time. In interacting multiparticle quantum systems, we expect this conservation to impose selection rules on permissible interactions. These selection rules should align with those derived from the theory's symmetries and conserved charges. However, unlike conventional conserved charges, the phase-space entropy of a system appears to be independent of any dimensionful quantities, underscoring its fundamental nature.

The close connection between quantum information, Riemannian geometry, and gravitation has appeared in multiple contexts, including the study of black hole horizons \cite{susskind2016computational, brown2018second} and optimal quantum circuits \cite{nielsen2005geometricapproachquantumcircuit, doi:10.1126/science.1121541}. Ultimately, this connection stems from the fact that any measurement inherently involves an interaction and, therefore, a transformation under the Poincar\'e group. We anticipate that extending the formalism developed here to incorporate relativistic effects and quantum fields will provide further insight into the interplay between quantum information and gravity. 

Continuous-variable quantum information, where information is encoded in continuous degrees of freedom satisfying the canonical commutation relations (e.g., the quadratures of the electromagnetic field), has found widespread application due to its versatility and ease of implementation \cite{braunstein2005quantum, weedbrook2012gaussian, adesso2014continuous}. For instance, continuous-variable quantum communication channels have enabled long-range key distribution in quantum cryptography \cite{zhang2024continuous}, with successful demonstrations over distances exceeding 200 km in fiber-optic links \cite{zhang2020long} and several commercial implementations. The security of these channels is rooted in entropic uncertainty relations, which bound the information accessible to an eavesdropper during an attack \cite{coles2017entropic}. Our work provides a quantitative framework for analyzing the effects of gravitational interactions and spatial geometry on continuous-variable quantum randomness, enabling us, for example, to incorporate geometric aspects when assessing the security of quantum key distribution. We anticipate that, as the range of quantum communication channels continues to increase or continuous-variable quantum information finds applications in space, such geometric considerations will become increasingly relevant.

Importantly, the phase-space entropy of a system can, in principle, be experimentally accessed through quantum state tomography \cite{smithey1993measurement} or weak measurements \cite{lundeen2011direct}, enabling direct experimentation in the laboratory. In our thought experiment with Alice, we assumed an idealized measuring apparatus capable of accessing individual microstates without altering the system's state. As a result, the microscopic phase-space entropy derived in this work is observer-independent. In real experiments and applications, however, observers only have access to macrostates, which depend on the measuring apparatus and the temporal order of the measurements. By coarse-graining the quantum phase space described in Section \ref{sec3}, we anticipate a complete quantum extension of the classical observational entropy introduced in \cite{vsafranek2020classical}, which would be the primary object of study in these settings.

\newpage

\appendix

\begin{widetext}
\section{Integrals}

\subsection{$\int_0^{\infty} e^{-r^2}L_n\left(2r^2\right)\log\left|L_n\left(2r^2\right)\right|\, rdr$}
\label{a1}

Using the close form and roots of the Laguerre polynomials, we can express
\begin{equation*}
\int_0^{\infty} e^{-r^2}L_n\left(2r^2\right)\log\left|L_n\left(2r^2\right)\right|\, rdr
=\frac{1}{2}\sum_{q=0}^n\sum_{\lambda\in\textrm{Roots}(L_n)}\left(\begin{matrix}n\\q\end{matrix}\right)\frac{(-2)^q}{q!}\int_0^\infty e^{-r^2} r^{2q+1}\log\left[\frac{(2r^2-\lambda)^2}{\lambda^2}\right]\,dr\nonumber
\end{equation*}
To carry out the integral on the right-hand side, we integrate by parts and make the change of variable $x=r^2-\frac{\lambda}{2}$,
\begin{multline*}
\int_0^\infty e^{-r^2} r^{2q+1}\log\left[\frac{(2r^2-\lambda)^2}{\lambda^2}\right]\,dr=4\int_0^\infty \frac{r\, \Gamma\left(q+1,r^2\right)}{2r^2-\lambda}\, dr\\
=2q!\sum_{l=0}^q\frac{1}{l!}\int_0^\infty\frac{r^{2l+1}e^{-r^2}}{r^2-\frac{\lambda}{2}}\, dr=q!\sum_{l=0}^q\frac{e^{-\frac{\lambda}{2}}}{l!}\int_{-\frac{\lambda}{2}}^\infty \frac{e^{-x}}{x} \left(x+\frac{\lambda}{2}\right)^ldx\nonumber
\end{multline*}
where $\Gamma(s,x)$ is the incomplete gamma function. Using the binomial formula,
\begin{multline*}
\int_{-\frac{\lambda}{2}}^\infty \frac{e^{-x}}{x} \left(x+\frac{\lambda}{2}\right)^ldx=\sum_{k=0}^l \int_{-\frac{\lambda}{2}}^\infty e^{-x}\left(\begin{matrix}l\\ k\end{matrix}\right)\left(\frac{\lambda}{2}\right)^{l-k}x^{k-1}\, dx\\
=-\left(\frac{\lambda}{2}\right)^l Ei\left(\frac{\lambda}{2}\right)+\sum_{k=1}^l\left(\begin{matrix}l\\ k\end{matrix}\right)\left(\frac{\lambda}{2}\right)^{l-k}\Gamma\left(k,-\frac{\lambda}{2}\right)\\
=-\left(\frac{\lambda}{2}\right)^l Ei\left(\frac{\lambda}{2}\right)+
e^{\frac{\lambda}{2}}\sum_{k=1}^l\sum_{s=0}^{k-1}\frac{l!\,(-1)^s}{s!\,(l-k)!\,k}\left(\frac{\lambda}{2}\right)^{l-k+s}\nonumber
\end{multline*}
where $Ei(x)$ is the exponential integral function. Putting all the pieces together,
\begin{multline}
\int_0^{\infty} e^{-r^2}L_n\left(2r^2\right)\log\left|L_n\left(2r^2\right)\right|\, rdr\\
=\sum_{q=0}^n\sum_{\lambda\in\textrm{Roots}(L_n)}\sum_{l=0}^q\left(\begin{matrix}n\\q\end{matrix}\right)\left[\frac{(-1)^{q+1}\lambda^l\, 2^{q-l-1}}{l!}e^{-\frac{\lambda}{2}}Ei\left(\frac{\lambda}{2}\right)+\sum_{k=1}^l\sum_{s=0}^{k-1}\frac{(-1)^{s+q}\,\lambda^{l-k+s}\,2^{q-l-s+k-1}}{s!\,(l-k)!\,k}\right]\label{int1}
\end{multline}

\subsection{$\int_C\frac{e^{-izp}dz}{(z-s)^j\, (z+s)^j\, (z-s^*)^j\, (z+s^*)^j}$}
\label{countour}

Let us consider the case $p<0$. The contour of integration is $C_{p<0}$ in Fig.~\ref{fig1},
\begin{equation*}
\int_{C_{p<0}}\frac{e^{-izp}dz}{(z-s)^j\, (z+s)^j\, (z-s^*)^j\, (z+s^*)^j}=2\pi i\left(\textrm{Res}_{-s}+\textrm{Res}_{s^*}\right)
\end{equation*}
where $\textrm{Res}_{-s}$ and $\textrm{Res}_{s^*}$ denote the residues of the integrand at $-s$ and $s^*$, respectively. To compute the residues, note that
\begin{equation*}
\textrm{Res}_s=\frac{e^{-izp}}{(j-1)!}\sum_{\ell=0}^{j-1}\left(\begin{matrix}j-1\\ \ell\end{matrix}\right)(-ip)^\ell \frac{d^{j-\ell-1}}{dz^{j-\ell-1}}\frac{1}{(z+s)^j\, (z-s^*)^j\, (z+s^*)^j}\ .
\end{equation*}
with
\begin{multline*}
\frac{d^k}{dz^k}\frac{1}{(z+s)^j\, (z-s^*)^j\, (z+s^*)^j}\\
=k!(-1)^k \sum_{\ell_1=0}^k\sum_{\ell_2=0}^{\ell_1}
\left(\begin{matrix}j+\ell_1-\ell_2-1\\ j-1\end{matrix}\right)\left(\begin{matrix}j+k-\ell_1-1\\ j-1\end{matrix}\right)\left(\begin{matrix}j+\ell_2-1\\ j-1\end{matrix}\right)
\frac{1}{(z+s)^{j+\ell_2}\, (z-s^*)^{j+\ell_1-\ell_2}\, (z+s^*)^{j+k-\ell_1}}
\end{multline*}
From this expression, we can easily obtain the expressions for the other residues by conjugating and/or changing the sign of $s$. Putting everything together,
\begin{multline*}
\int_{C_{p<0}}\frac{e^{-izp}dz}{(z-s)^j\, (z+s)^j\, (z-s^*)^j\, (z+s^*)^j}\\
=2\pi \sum_{\ell_3=0}^{j-1}\sum_{\ell_1=0}^{\ell_3}\sum_{\ell_2=0}^{\ell_1}\left(\begin{matrix}j+\ell_1-\ell_2-1\\ j-1\end{matrix}\right)\left(\begin{matrix}j+\ell_2-1\\ j-1\end{matrix}\right)\left(\begin{matrix}j+\ell_3-\ell_1-1\\ j-1\end{matrix}\right)\cdot\\
\cdot\frac{i^{-\ell_1+\ell_2-\ell_3}\,(-1)^{1+\ell_3}\,p^{j-1-\ell_3}}{2^{3j+\ell_3}\,(j-1-\ell_3)!}\, \frac{e^{-p\,\textrm{Im}(s)}\left[e^{ip\,\textrm{Re}(s)}\,s^*{}^{j+\ell_2}+(-1)^{\ell_1+\ell_2+\ell_3}\,e^{-ip\,\textrm{Re}(s)}\,s^{j+\ell_2}\right]}{[\textrm{Re}(s)]^{j+\ell_3-\ell_1}\,[\textrm{Im}(s)]^{j+\ell_1-\ell_2}\,|s|^{2(j+\ell_2)}}
\end{multline*}
The case $p>0$ can be computed similarly. Combining both cases into a single expression and rearranging the sums, we finally obtain
\begin{multline*}
\int_{C}\frac{e^{-izp}dz}{(z-s)^j\, (z+s)^j\, (z-s^*)^j\, (z+s^*)^j}\\
=2\pi\sum_{\ell_3=0}^{j-1}\sum_{\ell_2=0}^{\ell_3}\sum_{\ell_1=\ell_2}^{\ell_3}\left(\begin{matrix}j+\ell_3-\ell_1-1\\ j-1\end{matrix}\right)\left(\begin{matrix}j+\ell_2-1\\ j-1\end{matrix}\right)\left(\begin{matrix}j+\ell_1-\ell_2-1\\ j-1\end{matrix}\right)\frac{e^{-2|p|R}}{(j-1-\ell_3)!}\cdot \\
\cdot\frac{(-i)^{\ell_1}\,R^{\ell_1-j-\ell_3}\,|p|^{j-1-\ell_3}}{4^{3j+\ell_3}\, x^{j+\ell_2}\,(R^2+x^2)^{j+\ell_1-\ell_2}}\,\left[e^{-2i|p|x}\,(x+iR)^{j+\ell_1-\ell_2}+(-1)^{\ell_1}\,e^{2i|p|x}\,(x-iR)^{j+\ell_1-\ell_2}\right]
\end{multline*}
\end{widetext}

\bibliography{apssamp}

\end{document}